# Analysis of Proton Radiography Images of Shock Melted/Damaged Tin

Hanna Makaruk[1], Nikita A. Sakhanenko [1,2], David B. Holtkamp[1], Tiffany Hayes[1,3]
Joysree Aubrey[1]

[1]Physics Division, Los Alamos National Laboratory
[2] Computer Science Dept., University of New Mexico
[3] Physics and Astronomy Dept., University of New Mexico

## Abstract
     Tin coupons were shock damaged/melted under identical conditions with a diverging high explosive shock wave. Proton Radiography images and velocimetry data from experiments with seven different tin coupons of varying thickness are analyzed. Comparing experiments with identical samples allowed us to distinguish between repeatable and random features. Shapes and velocities of the main fragments are deterministic functions of the coupon thickness; random differences exist only at a small scale. Velocities of the leading layer and of the main fragment differ by the same value independently of coupon thicknesses, which is likely related to the separation energy of metal layers.

## Experiment

     Dynamic phase change in shocked materials is one of the most challenging topics in equation-of-state research. The physical conditions are extreme, resembling rather late stages of stellar evolution than typical engineering processes. When pressure increases sufficiently, metals can undergo a solid-liquid phase transition. This article considers such experiments with tin [1].

     All experiments considered here have the same axially symmetric setup. Tin coupons 5.1 cm in diameter and 4.76 mm - 12.7 mm thick are placed on a high explosive (HE) 12.7 mm thick disk also 5.1 cm in diameter. A point detonator is glued to the opposite side of the HE in the center. The experimental data obtained are Proton Radiography (PRAD) [2,3,4] image series and 1-D signals from Velocity Interferometer System for Any Reflector (VISAR). VISAR records the coupon's external surface velocity [5] by measuring the Doppler shift of the 532 nm laser light. The reflected, Doppler shifted beam is split, delayed, and interfered with itself. This produces a signal (i.e. phase) related to the acceleration of the observed surface. This produces a very precise measurement of the reflective surface velocity. The main limitation of this type of velocity measurement is that only surfaces possessing good reflective quality can be measured this way. Internal material and density fluctuations inside a disintegrating/melting coupon cannot be observed in this way. During an interval of 10's of microseconds after detonation, a series of PRAD images are obtained using a proton beam orthogonal to the axis of symmetry. A proton beam is used instead of X-rays to ensure proper penetration into a relatively dense metal coupon. Complexity and cost limits the number of experiments and increases the need for extracting quantitative information with enhanced accuracy from PRAD images.

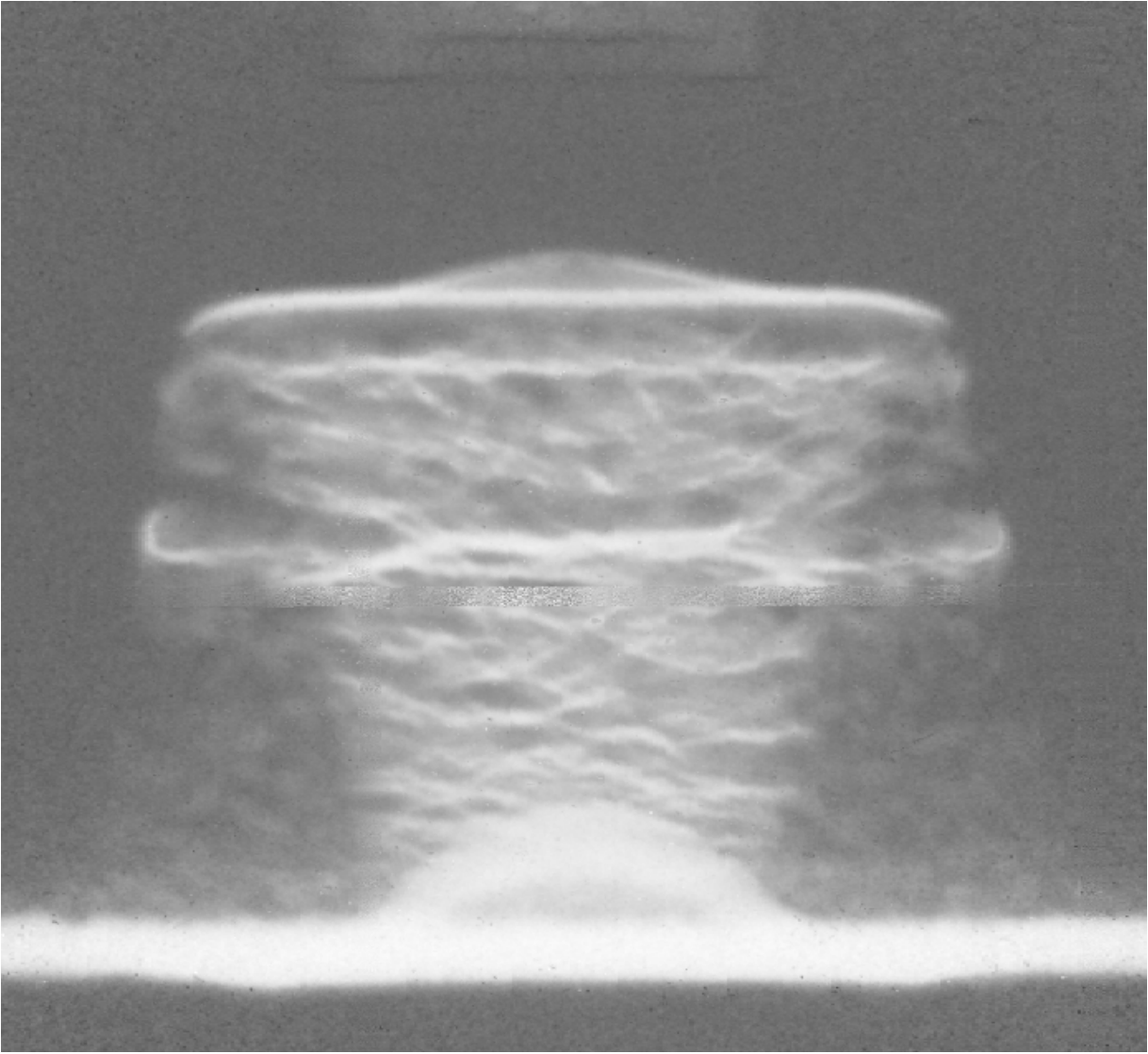

Fig. 1 Typical PRAD image of (initially) a 12.7 mm thickness tin coupon with visible fragmentation after shock acceleration.

## Deterministic vs. Stochastic

Knowing which features of the system's evolution are deterministic and which are stochastic is important for understanding the physical process and for quantitative comparison of experiments with hydrocode modeling. To establish these information three experiments with coupon thicknesses 3.175 mm, 8 mm, and 12.7 mm were repeated.

Each pixel of a PRAD image represents a 0.1 x 0.1 mm$^2$ square of a 2D radiogram of the 3D distribution of the material. The radiogram is made by a parallel proton beam. Overlapping and then subtracting images from the same stage of the damage/melting evolution in repeated experiments provides qualitative analysis of the process. All the main axially symmetric features of the system of compared experiments are the same. Their location, velocities, and macroscopic shape evolution are also the same.

These features are deterministic within 1 mm precision in shape changes and less than 1.5% in leading surface velocity. Thus, these features should be reproducible in a model or a numerical simulation. On the other hand, all visible non-axially symmetric structures, in particular shapes and locations of density fluctuations in the central area vary stochastically from experiment to experiment. Only the average properties of these fluctuations (size, number) are reproducible.

A method to quantify image-to-image comparison, presented here, as a measure of distances between the overlapped images, is required to enhance accuracy of this comparison. Since large differences between color values of corresponding pixels indicate low overlapping areas, we propose an $l^2$ measure. More important, this imagery technique introduces only small value fluctuations for all the pixels.

$$D(I_1 I_2) = 1/2 \sqrt{\sum_{k=1}^{X} \sum_{l=1}^{Y} \frac{(C_1(i,k) - C_2(i,k))^2}{X \cdot Y}}$$

In this measure $D(I_1,I_2)$ is a distance between images $I_1$ and $I_2$, $C_1(i,k)$ and $C_2(i,k)$ are color values of their pixels with coordinates $(i,k)$, and $XY$ is a scaling factor. This measurement returns 255 as the maximum possible distance between images (a black square minus a white square), whereas the distance of an image to itself is 0. The difference between two identical static PRAD images measures $\leq 0.5$, while typically the distance between two consecutive images from the same experiment is 15-20. Correspondingly, the distance between overlapping images from a pair of repeated experiments with a thick coupon, when non-axially symmetric density fluctuations are observed, is around 10. When no such fluctuations are visible, the distance is around 2 for the same stage of a pair of experiments with a thin coupon. This shows that experiments are repeatable with high accuracy.

**Contour Detection and Velocity Measurements**

Tracking shape evolution of metal fragments and their movement requires precise contour detection. Since 5 cm diameter tin coupons strongly attenuate the 800 MeV protons of PRAD, many of the images have low contrast. Additionally, a PRAD-specific overshot-undershot limbing artifact [6] is present in the images (Fig. 2).

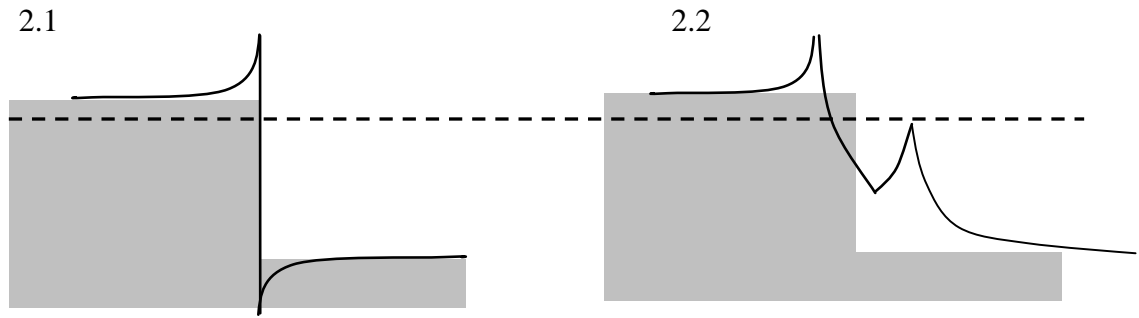

Fig. 2. Gray step represents material density change imaged by PRAD. 2.1 Resulted artifact when magnetic lenses collimating the proton beam are perfectly focused. 2.2 The artifact when magnetic lenses are not perfectly focused. Dashed line represents a possible cut-off level that would produce a unique continuous contour in both cases.

Gradient-based contour detection methods, standard in contemporary edge detection, are applicable only in case 2.1 (Fig. 2), but not in 2.2 where significant chromatic "blurring" occurs because of the energy loss of the protons passing through the material. In case 2.2 a gradient-based method may not detect any contour or it may detect two contours. Frequently, all situations intertwine making a proper detection of contour's coordinates difficult. Although the contours from the analyzed images were shifted from the true location of the density edge by a fraction of a millimeter, the resulting error is negligible when the shift is consistent for all compared contours from a series.

A level-set-type method [7] that we developed for PRAD images works in both cases 2.1 and 2.2. The method starts from choosing a threshold value. Every pixel of an image darker than the value is converted to black; otherwise, it becomes white. Eroding the black area one pixel deep and subtracting it from the un-eroded image results in a single continuous contour, each point of which is saved as coordinates (x, y). A fragment's velocity is calculated from the contours extracted from a single image series. For each pair of consecutive images, a difference $\Delta y$ along y-axis is calculated, for each x-coordinate. $v_y(x_i) = \Delta y/\Delta t$ is an instantaneous vertical velocity for x-coordinate $x_i$ and time difference $\Delta t$ between $t_k$ and $t_{k+1}$ of the $k$, and $k+1$ image. Instantaneous velocities of a fragment are then averaged in time and, if appropriate, along $x$, since the fragment velocities are constant after the explosion.

The velocities of the central part of the leading surface that are obtained this way are compared with appropriate velocity measurements provided by VISAR. Repeating this process for all the experiments appears to validate the technique. VISAR velocimetry is well established. However, it can only measure the velocity of a surface directly exposed to the probe light. For the best results the surface should be reflecting, which typically requires a solid phase. On the other hand, velocity measurements from images can be done for any fragment with visible edges regardless of its phase. Moreover, the fragment can be in any location, even those occluded by other fragments. Precision may vary if an edge is less visible.

## Velocities of different parts of the system

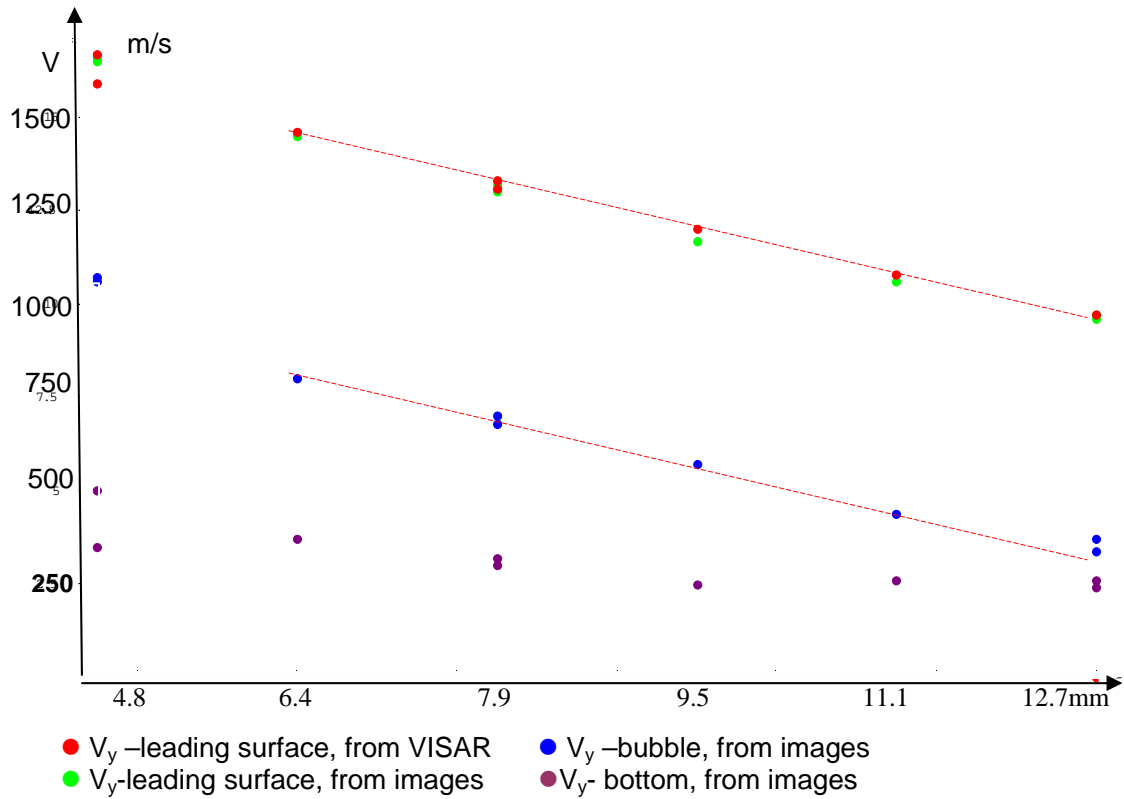

- ● $V_y$ –leading surface, from VISAR
- ● $V_y$ –bubble, from images
- ● $V_y$-leading surface, from images
- ● $V_y$- bottom, from images

-Trend connecting the leading surface velocities and the same trend moved 612m/s, connecting velocities of the buble.
Fig. 3 Velocities of different parts of the system as a function of sample thickness

After measuring the velocities of fragments with the method described above, a graph representing these velocities as a function of sample thickness was plotted. Velocities measured by visar were superimposed on this figure (fig 3). Analyzing these data, we observed a correlation: the velocity of a bubble (a hemispherical structure in the lower part, containing most of the material) is systematically 612 m/s smaller than the leading surface velocity, i.e. within 5% difference in all the examined coupons of tin. The correlation is unlikely to be a coincidence; rather, it reflects the difference in tension produced by the reflected HE Taylor wave until it reaches a level that is low enough to be inadequate to cavitate melted tin. Thus, the residual melted bubble remains intact.

Additionally, for coupon thicknesses 6.14mm – 12.7 mm the relation between velocity of the leading surface and coupon thickness is linear. The relation between velocity of the bubble surface and coupon thickness is, in consequence of these two observations, also linear and has the same coefficient as for the leading surface. The 4.76 mm coupon has the same 612 m/s difference in velocity between the leading surface and the main fragment (bubble) but all velocities are higher than it would be for the liner correlation visible for all the thicker fragments. The difference was in this sample being completely melted after the explosion, opposite to the thicker ones.

**Error Bars & Stochastic Fluctuations**

For each pair of repeated experiments observed differences between images are of similar magnitude. Unfortunately, it is not obvious how much of the differences are caused by measurement errors and how much by genuine stochastic fluctuations in the system. These differences provide an upper estimate for the measurement error and for the stochastic fluctuations. There were individual cases in which stochastic fluctuations are the most probable reason for differences between data from repeated experiments, e.g., inhomogeneities in trunk areas that differ in location and size, or fluctuations visible on bubble and bottom surfaces. Differences in leading surface velocities were smaller than 1.5%, and even smaller from VISAR measurements - the most precise technique currently used. For a "bubble" fragment, that is partially obscured by other fragments, the value was <3% of this fragment velocity.

The method presented here is applicable to PRad images of significantly lower contrast and quality than the classic contour detection methods.